\documentclass{aastex631}

\usepackage{graphicx}
\graphicspath{{./}{/Users/stary/Desktop/tex/M31}}
\DeclareGraphicsExtensions{.pdf,.jpg,.png}
\usepackage{float}

\usepackage{threeparttable}

\begin{document}

\title{Searching for Radio Outflows from M31* with VLBI Observations}
\email{sijiapeng@smail.nju.edu.cn}
\email{lizy@nju.edu.cn}

\author[0000-0001-8492-892X]{Sijia Peng}
\affiliation{School of Astronomy and Space Science, Nanjing University, Nanjing 210023, People's Republic of China}
\affiliation{Key Laboratory of Modern Astronomy and Astrophysics, Nanjing University, Nanjing 210023, People's Republic of China}
\affiliation{Shanghai Astronomical Observatory, Chinese Academy of Sciences, Shanghai 200030, People's Republic of China}

\author[0000-0003-0355-6437]{Zhiyuan Li}
\affiliation{School of Astronomy and Space Science, Nanjing University, Nanjing 210023, People's Republic of China}
\affiliation{Key Laboratory of Modern Astronomy and Astrophysics, Nanjing University, Nanjing 210023, People's Republic of China}

\author[0000-0003-3096-3062]{Lor\'{a}nt O. Sjouwerman}
\affiliation{National Radio Astronomy Observatory, Socorro, NM 87801, USA}

\author[0000-0001-7254-219X]{Yang Yang}
\affiliation{School of Astronomy and Space Science, Nanjing University, Nanjing 210023, People's Republic of China}
\affiliation{Purple Mountain Observatory, Chinese Academy of Sciences, Nanjing 210033, People's Republic of China}

\author[0000-0001-7369-3539]{Wu Jiang}
\affiliation{Shanghai Astronomical Observatory, Chinese Academy of Sciences, Shanghai 200030, People's Republic of China}
\affiliation{Key Laboratory of Radio Astronomy, Chinese Academy of Sciences, Shanghai 200030, People's Republic of China}

\author[0000-0003-3540-8746]{Zhi-Qiang Shen}
\affiliation{Shanghai Astronomical Observatory, Chinese Academy of Sciences, Shanghai 200030, People's Republic of China}
\affiliation{Key Laboratory of Radio Astronomy, Chinese Academy of Sciences, Shanghai 200030, People's Republic of China}





\defcitealias{2017ApJ...845..140Y}{Y17}

\begin{abstract}
As one of the nearest and most dormant supermassive black holes (SMBHs), M31* provides a rare but promising opportunity for studying the physics of black hole accretion and feedback at the quiescent state. Previous Karl G. Jansky Very Large Array (VLA) observations with an arcsec resolution have detected M31* as a compact radio source over centimeter wavelengths, but the steep radio spectrum suggests optically-thin synchrotron radiation from an outflow driven by a hot accretion flow onto the SMBH. 
Aiming to probe the putative radio outflow, we have conducted milli-arcsec-resolution very long baseline interferometric (VLBI) observations of M31* in 2016, primarily at 5 GHz and combining the Very Long Baseline Array, Tianma-65m and Shanghai-25m Radio Telescopes.
Despite the unprecedented simultaneous resolution and sensitivity achieved, no significant ($\gtrsim 3\sigma$) signal is detected at the putative position of M31* given an RMS level of $\rm 5.9~\mu Jy\ beam^{-1}$, thus ruling out a point-like source with a peak flux density comparable to that ($\sim30~\mu Jy\ beam^{-1}$) measured by the VLA observations taken in 2012.
We disfavor the possibility that M31* has substantially faded since 2012, in view that a 2017 VLA observation successfully detected M31* at a historically-high peak flux density ($\sim75~\mu Jy\ beam^{-1}$ at 6 GHz).
Instead, the non-detection of the VLBI observations is best interpreted as the arcsec-scale core being resolved out at the milli-arcsec-scale, suggesting an intrinsic size of M31* at 5 GHz larger than $\sim300$ times the Schwarzschild radius. Such extended radio emission may originate from a hot wind driven by the weakly accreting SMBH.
\end{abstract}




\section{Introduction} 
For most of their lifetime, supermassive black holes (SMBHs), commonly residing in galactic nuclei, obtain mass from the ambient gas at a rate well below the Eddington limit \citep{1982MNRAS.200..115S,2002MNRAS.335..965Y,2004MNRAS.351..169M}, which is generally thought to be mediated by a radiatively inefficient, hot accretion flow \citep{2014ARA&A..52..529Y}. 
As such, most SMBHs in the local Universe manifest themselves as low-luminosity active galactic nuclei (LLAGNs; \citealp{2008ARA&A..46..475H}).
Direct probes of these LLAGNs prove to be challenging and generally require high-resolution, high-sensitivity observations. Nevertheless, studies of LLAGNs are crucial for our comprehensive understanding of SMBH accretion and feedback over cosmic time \citep{2012ARA&A..50..455F,2013ARA&A..51..511K}.

It is now widely accepted that the hot accretion flow is symbiotic with outflows in the form of relativistic jets and/or hot winds (see review by \citealp{2014ARA&A..52..529Y}). The prevalence of jets has long been established from radio interferometric surveys of nearby LLAGNs, in which synchrotron cores, with or without elongated components, are frequently detected and conventionally interpreted as a highly collimated and magnetized relativistic outflow \citep[e.g.,][]{2000ApJ...542..186N,2001ApJS..133...77H}. 
These jets can inject an enormous amount of mechanical energy and momentum into the environment, providing the so-called radio-mode or kinetic-mode feedback.
Compelling evidence for radio-mode feedback comes from the observation of radio bubbles (often spatially coincident with X-ray cavities) inflated by relativistic jets, which are typically found in massive elliptical galaxies, galaxy groups, and galaxy clusters \citep{2012NJPh...14e5023M}.
The hot wind, on the other hand, is a generic prediction of both theories \citep{1999MNRAS.303L...1B} and numerical simulations \citep{2012ApJ...761..130Y,2012MNRAS.426.3241N}.
Originating from over a wide radial range in the hot accretion flow, the wind may affect the accretion process of the black hole \citep{2014ARA&A..52..529Y}. 
Moreover, the wind has a much larger opening angle compared to the jet, which facilitates the coupling of its momentum and kinetic energy to the ambient gas \citep{2014ARA&A..52..529Y}. 
For these reasons, recent years have witnessed a growing interest in the hot wind as an efficient means of kinetic feedback, in addition to the conventional jet-driven feedback.  
In particular, in the influential cosmological simulations of galaxy formation and evolution, IllustrisTNG \citep{2017MNRAS.465.3291W,2018MNRAS.473.4077P}, a kinetic feedback mode mimicking an isotropic wind from weakly accreting SMBHs is invoked to quench star formation in intermediate- to high-mass galaxies.
On the observational side, however, 
direct evidence for LLAGN-driven hot winds is still limited \citep{2021NatAs...5..928S,2022ApJ...926..209S}.
Moreover, it remains unclear whether and how an efficient kinetic feedback is materialized by the most dormant SMBHs, such as the one hosted by our own Galaxy, commonly known as Sgr A*.  
Indeed, it remains an open question whether Sgr A* produces a jet (see \citealp{2013ApJ...779..154L,2019ApJ...875...44Z}, and discussions therein).

Clearly, our understanding of the physics of LLAGNs will not be complete without a census of SMBHs at the most quiescent state.
Two primary factors make the SMBH in M31, also known as M31*, a prime target: i)  it is the second nearest SMBH (next to Sgr A*), allowing for a close-up view of the accretion flow and the putative outflow (jet and/or wind),
and ii) among known LLAGNs, M31* shows an exceptionally low Eddington ratio ($\rm \sim 10^{-8}$) comparable to that of Sgr A*. 
M31* has been detected only in the radio and X-rays to date. \citet{1992ApJ...390L...9C} first identified a compact radio source coincident with the nucleus of M31, based on Very Large Array (VLA) observations at a frequency of 8.4 GHz.
Its physical association with the SMBH was reinforced by a follow-up 8.4 GHz observation that found mild variability ($\rm \sim 30-40\ {\mu}Jy$; \citealp{1993ApJ...417L..61C}). 
\citet{2010ApJ...710..755G} measured an average flux density of $\rm \sim 50\ {\mu}Jy$ at 4.9 GHz based on VLA observations taken between 2002--2005.
The X-ray counterpart of M31* was not firmly detected until a much later time by sensitive Chandra observations \citep{2010ApJ...710..755G,2011ApJ...728L..10L}. 
Intriguingly, M31* has exhibited X-ray flares since early 2006, with amplitudes similar to those seen in Sgr A* \citep{2011ApJ...728L..10L}.

More recently, \citet[][hereafter \citetalias{2017ApJ...845..140Y}]{2017ApJ...845..140Y} presented the results from a multi-epoch, multi-frequency VLA observing campaign of M31* conducted between 2011--2012. 
Based on these observations of unprecedented sensitivity ($\sim \rm 2.0~\mu Jy~beam^{-1}$ in the X [8--12 GHz], Ku [12--18 GHz], and K [18--27 GHz] bands), \citetalias{2017ApJ...845..140Y} detected M31* at 10, 15 and 20 GHz for     the first time, and measured the quasi-simultaneous spectral index $\alpha \approx -0.45 $ ($S_\nu \propto \nu^{\alpha}$) over 5--20 GHz. 
This steeper spectral index is markedly different from that of Sgr A* ($\alpha \sim 0.3$) over the centimeter-to-millimeter range. 
The flat radio spectrum of the latter is generally thought to be the signature of the self-absorbed synchrotron core or the base of the jet \citep{2010LNP...794..143M}. 
In this regard, the steep spectral index of M31* may be understood as arising from the low-density, optically-thin part of a putative outflow, presumably located at distances well beyond the black hole event horizon.
However, the bulk of the radio flux of M31* comes from a compact core, which is unresolved even under the resolution of $ 0\farcs2$ (0.76 pc) \citepalias{2017ApJ...845..140Y}.
Only in the highest sensitivity image at 6 GHz faint emission on a scale of $1\arcsec - 2\arcsec$ (3.8 -- 7.6 pc) was revealed, which might be tracing the putative radio outflow \citepalias{2017ApJ...845..140Y}.

Motivated by the findings from the VLA campaign, a natural next step would be to probe substructures in the compact radio core of M31* at smaller physical scales. 
In this work, we report our effort in searching for the $mas$-scale radio emission from a putative outflow driven by M31*, based on Very Long Baseline interferometry (VLBI) observations,
for the first time, achieving a sensitivity of $\sim 10~\mu$Jy beam$^{-1}$ simultaneously with an angular resolution of $\sim$ 1 mas. 
In Section \ref{sec:obs}, we describe the VLBI observations and complementary data from VLA, as well as our data reduction procedures. 
The resultant VLBI images are presented in Section \ref{sec:result}, which do not reveal significant signals. 
Implications of the non-detection are discussed in Section~\ref{sec:dis}. For the distance of M31, $D \approx$ 780 kpc \citep{1998ApJ...503L.131S}, and the SMBH's dynamical mass of $\rm \sim 1.4\times10^8~{\rm M}_{\odot}$ \citep{2005ApJ...631..280B},  1 $\rm mas$ corresponds to $\sim 0.0038$ pc or $\sim 280\ R_{\rm Sch}$, where $R_{\rm Sch}$ ($\rm = 2GM_{BH}/c^2$) is the Schwarzschild radius.

\section{VLBI Observations and Complementary VLA Data}
\label{sec:obs}
We have carried out VLBI observations of M31* in four epochs in 2016 (Program ID: BL223; PI: Z. Li), each with an integration time of 7-hr ($\sim $ 4.5 hr on-source) at $\sim$5 GHz/C-band (Table~\ref{tab:infor}). 
Our observations were mainly built on the Very Long Baseline Array (VLBA), which consists of ten 25-meter antennas in the United States.  
Meanwhile, we attempted to include the Tianma 65-meter radio telescope (T6) and Shanghai 25-meter radio telescope (SH) in Shanghai, China, to enhance both the sensitivity and resolution in all four epochs. These were among the first correlated observations between VLBA and T6 at 5 GHz.  
In the first epoch, the central frequency was set at 5.9 GHz. 
Unfortunately, we failed to obtain fringes for T6 in the first epoch (BL223\_A) due to a malfunction of the atomic clock. 
Therefore, the first epoch involved only the VLBA stations. Adding SH station to check fringes, 
the central frequency was then switched to 4.9 GHz. 
Fringes were successfully obtained between VLBA, T6, and SH in the next three epochs (BL223\_B, C, and D).
It is noteworthy that in epoch BL223\_B, two VLBA antennas, Saint Croix (SC) and Pie Town (PT), did not participate;  
in epoch BL223\_D, SH did not participate and there was no data recorded in Hancock (HN).
In all four epochs, the observations were conducted with a bandwidth of 256 MHz per polarization and a data rate of 2048 Mbps.
The observing phase center was set to be [$\rm RA , DEC$] = $[00^{h}42^{m}44.32^{s}, +41\arcdeg16\arcmin08.50\arcsec]$ (J2000), the nominal central position of M31 provided by the NASA/IPAC Extragalactic Database.
We note that this position is offset by 103 mas from the radio centroid of M31*, [$\rm RA , DEC$] = $[00^{h}42^{m}44.325^{s}, +41^{o}16'08.43'']$\footnote{We recommend future radio observations of M31* to use this position.}, which was measured from the VLA 6.0 GHz/C-band observations of \citetalias{2017ApJ...845..140Y}.
J0038+4137 was adopted for phase-referencing at an angular separation of $ \rm 1.14^{o}$, with a cycle time of 290 seconds including 196 seconds on M31*, 64 seconds on J0038+4137, and 30 seconds of switching time.
J0136+4751 served as the flux calibrator and fringe-finder, observed in two separate scans each for seven minutes in each epoch.

\begin{deluxetable*}{ccccccc}
\tablecaption{Log of VLBI Observations\label{tab:infor}}
\tablecolumns{7}
\tablenum{1}
\tablewidth{0pt}
\tablehead{
\colhead{ Epoch} & \colhead{Date} & \colhead {Frequency} & \colhead{Time} &
\colhead{Antennas} &
\colhead{ RMS} & \colhead{Synthesized beam} \\
\colhead{} & \colhead{(yyyy-mm-dd)} & \colhead{(GHz)} & \colhead{(hours)} & \colhead{} &\colhead{$(\rm \mu Jy\ beam^{-1})$}  & \colhead{($\rm mas \times mas,\ degree$)}
}
\colnumbers
\startdata
BL223\_A & 2016-04-20 & 5.9 & 4.5 & VLBA & 16.0  & $\rm 2.31\ \times1.27, -34.4$ \\
BL223\_B & 2016-08-29 & 4.9 & 4.5 & T6 + SH + VLBA - PT - SC & 10.9 & $\rm 1.85\ \times0.74, -6.4$ \\
BL223\_C & 2016-10-4 & 4.9 & 4.5 & T6 + SH + VLBA  & 10.0  & $\rm 1.98\ \times0.72, -2.7$ \\
BL223\_D & 2016-11-18 & 4.9 & 4.5 & T6 + VLBA - HN & 9.0  & $ \rm 1.88\ \times0.71, -4.4$ \\
Combine\_BCD & - &  4.9 & 13.5 & -  & 5.9 & $\rm 1.90\ \times0.72, -6.7$
\enddata

\tablecomments{(1) Name of Epochs. (2) Observational date. (3) Central frequency. (4) The on-source integration time. (5)Antennas that participated in each observing epoch. `VLBA'means that all 10 VLBA stations participated in a specific epoch. T6 represents Tianma-65m radio telescope (China) and SH represents Shanghai-25m radio telescope (China).
``+'' means add, while ``-'' means exclude. For example, ``T6 + SH + VLBA - PT - SC'' means that the antennas included in the observation were Tianma-65m (T6), Shanghai-25m (SH) and the VLBA antennas excluding Pie Town (PT) and Saint Croix (SC). 
(6) Map RMS level measured from Difmap. (7) Synthesized beam FWHM of major and minor axes, and position angle of the major axis measured from Difmap.} 
\end{deluxetable*}

The data were calibrated using the Astronomical Image Processing System\footnote{http://www.aips.nrao.edu/index.shtml} \citep[AIPS, version: 31DEC16,][]{1996ASPC..101...37V} in standard VLBI data processing\footnote{http://www.aips.nrao.edu/TEXT/PUBL/COOK9.PS.gz} and imaged in Difmap\footnote{https://www.cv.nrao.edu/adass/adassVI/shepherdm.html} \citep{1994BAAS...26..987S} referencing the steps in the Difmap cookbook\footnote{ http://ftp.ira.inaf.it/Computing/manuals/difmap/cookbook.ps}. We first corrected sampler biases using task ACCOR, and then used the flux calibrator to align the phases and delay offsets among all baseband channels by task FRING. Global-fringe solutions (task FRING) were obtained from both the phase and flux calibrators. After that, we set INTERPOL=`AMBG' in task CLCAL to obtain the linear phase interpolation for the target source (M31*) from the phase calibrator J0038+4137.
Then, amplitude correction (task APCAL) was derived from the system temperature and the gain curves of stations. Only amplitude bandpass corrections (task BPASS) were applied based on flux calibrator scans. 
Finally, we extracted the target visibilities into a single file and further concatenated the visibilities of the last three epochs, which are of the same central frequency of 4.9 GHz, to obtain a combined one called $\rm Combined\_BCD$ hereafter. 
As for imaging, maps were produced in Difmap by setting a cellsize of 0.2 mas with a natural weighting.
The image noise level (RMS) and synthesized beam sizes of the maps are listed in Table \ref{tab:infor}. 
The RMS is 16.0, 10.9, 10.0, and 9.0 $\rm \mu Jy\ beam^{-1}$ for epochs A, B, C, and D, respectively. These RMS values are somewhat higher than the theoretical RMS of 6.4 $\rm \mu Jy\ beam^{-1}$, estimated from the EVN calculator\footnote{http://www.evlbi.org/cgi-bin/EVNcalc} for an on-source time of 4.5 hr combining VLBA, T6, and SH. 
This is understood because 
the co-view integration time of the easternmost (SC) and westernmost (T6) telescopes was less than 3 hr.
The $\rm Combined\_BCD$ image has an RMS of 5.9 $\rm \mu Jy\ beam^{-1}$ and a beam size of 1.90 mas$\times$0.72 mas. It is also noteworthy that we have successfully fringe-detected and imaged the phase calibrator J0038+4137. 
A comparison with previous VLBA survey data\citep{2016AJ....152...12L} finds excellent morphological agreement in the core-jet structure of J0038+4137, which verifies our observational setup and data calibration procedure.

We supplemented our VLBI observations with an archival VLA observation of M31*, which was performed on November 11, 2017 (Project ID:17B-148; PI: A. Brunthaler). The observation was taken at the 6.0 GHz/C-band (with a bandwidth of 4 GHz) for an integration time of 6 hr under B-configuration. 
This was the first VLA C-band observation of M31* since the monitoring campaign of \citetalias{2017ApJ...845..140Y}, which helps constrain the flux density of M31* at a time close to our VLBI observations. 
We used the Common Astronomy Software Applications package (CASA) VLA pipeline\footnote{https://science.nrao.edu/facilities/vla/data-processing/pipeline/\#section-7} \citep[pipeline version: 6.2.1,][]{2022PASP..134k4501C} to calibrate the data, applying additional manual flagging. The CASA pipeline performed automated radio frequency interference flagging. Antenna position corrections, flux density scaling, delay, bandpass, gain, and phase calibrations were applied. We also examined the diagnostic information and plots generated by the VLA pipeline.
A cleaned image was manually obtained with the task `TCLEAN' by setting the `MTMFS' mode with NTERMS=2, GRIDDER=`STANDARD', CEll=`0.2ARCSEC', IMSIZE=[1000, 1000], and natural weighting. The image has an RMS of $\rm 5.4\ \mu Jy\ beam^{-1}$ and a synthesized beam of $\rm 1\farcs25 \times 1\farcs22$ ($\rm 46.4\arcdeg$).
\\

\section{Results}
\label{sec:result}
\begin{figure}
\centering
\includegraphics[width=1\textwidth]{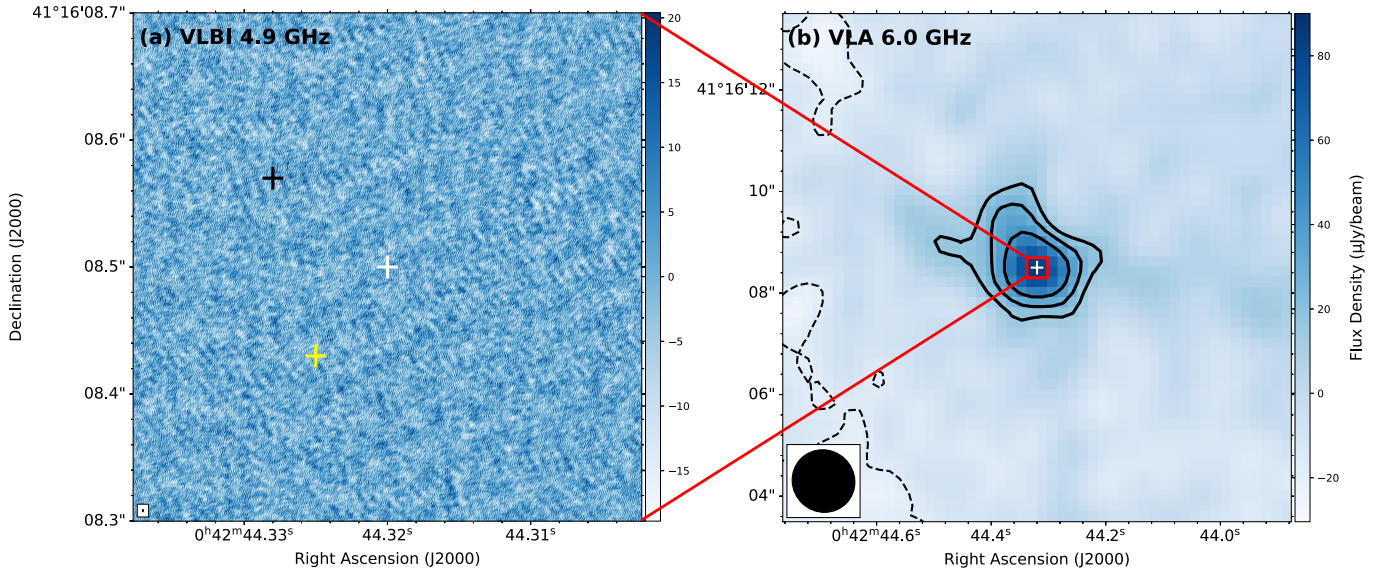}
\caption{(a) The map of concatenated data which combines the visibilities in epochs B, C, and D\label{fig:combined}. The image has a size of $\rm 400\ mas \times 400\ mas$ centering at the VLBI phase center marked by the white cross. The yellow and  black crosses mark the centroid position of M31* from the VLA 6.0 GHz/C-band observations of \citet{2017ApJ...845..140Y} at [$\rm RA , DEC$] = $[00^{h}42^{m}44.325^{s}, +41^{o}16'08.43'']$, and from the 2017 VLA observation at [$\rm RA , DEC$] = $[00^{h}42^{m}44.328^{s}, +41\arcdeg16\arcmin08.57\arcsec]$, respectively. The black ellipse at the lower left corner denotes the VLBI synthesized beam size. (b) The clean map of 2017 VLA 6.0 GHz/C-band observation: The black contours are at levels of $\rm (-3, 3, 5, 8) \times RMS$ (5.4 $\rm \mu Jy\ beam^{-1}$), with the negative components shown as dashed lines. The black ellipse at the lower left corner denotes the VLA synthesized beam size.} 
\end{figure}

We do not detect any emission from a point-like source at the location of M31* above a significance of 3$\sigma$ in any of the individual VLBI epochs.
In any epoch, no point-like source at a significance of $\gtrsim$ 3$\sigma$ can be found at the imaging phase center or the putative radio position of M31*, in contrast to the highly compact core of Sgr A*.  
There is neither any appreciable signal within the central $\rm 400 \times 400~mas$ nor $\rm 1.5 \times 1.5~pc$, a region comparable to the synthesized beam size of the VLA A-array 6.0 GHz image of \citetalias{2017ApJ...845..140Y}, in which significant emission was detected.   
The same situation holds in the $\rm Combined\_BCD$ map,
shown in Figure \ref{fig:combined}, which has an RMS ($\rm 5.9\ \mu Jy\ beam^{-1}$) at least 1.5 times lower than in the individual epochs.

The detectable largest angular scale\footnote{The largest angular scale (LAS) is defined as $\rm LAS \approx 2063\ \lambda / (2*B_{min})\ $arcsec, where $\lambda$ denotes the observational wavelength ($\rm cm$) and $B_{min}$ denotes the minimum length of baselines ($\rm m$) on the $uv$-plane.} (LAS) of the VLBI observations is $\sim \rm 2\arcsec$, corresponding to 7.6 pc at the distance of M31, which is determined by the shortest baseline between T6 and SH (6.1 km).
Since the shortest baseline is comparable to the length of the VLA baselines, we check the visibility between this baseline in AIPS to find any possible signal. 
However, no significant signal is detected due to a much-reduced sensitivity of $\rm 0.25\ mJy\ beam^{-1}$ with this baseline alone. We have also inspected the map that contains only short VLBA baselines ($\sim$200--2400 km) afforded by the telescopes in  Pie Tow (PT), Owens Valley (OV), Fort Davis (FD), Kitt Peak (KP), Los Alamos (LA), and Brewster (BR), whose LAS is $\sim \rm 0.2\arcsec$, but no significant signal is detected, either. The image combining these six antennas alone has a sensitivity of $\rm 11.0\ \mu Jy\ beam^{-1}$ and a synthesized beam size of $ \rm 10.3\ mas \times4.3\ mas$ (with a position angle of 34.7$\arcdeg$).
Combining the visibility of all four epochs has been tried as well though the frequencies are slightly different, again resulting in non-detection. 

In contrast, a nuclear source, presumably M31*, was clearly detected in the 2017 VLA 6.0 GHz observation, as illustrated in Figure \ref{fig:combined}b. 
The overall morphology of the nuclear source is highly consistent with that seen in the VLA 6.0 GHz observations of \citetalias{2017ApJ...845..140Y}, i.e., a compact core dominating the observed flux, plus weak ``plumes'' reaching out from the core to a projected distance of $2\arcsec$, most prominently toward the northeast. Fitting the nuclear source detected using the VLA in 2017 with an elliptical Gaussian model by the CASA task `IMFIT', we find a peak flux density of $\rm 74.4\pm5.8\ {\mu}Jy$ and an integrated flux density of $\rm 116\pm 14\ {\mu}Jy$, with a deconvolved size of $\rm 1.29\arcsec \times 0.52\arcsec$ (with a position angle of $\rm 44\arcdeg$) at a centroid position of [$\rm RA , DEC$] = $[00^{h}42^{m}44.328^{s}, +41\arcdeg16\arcmin08.57\arcsec]$ (marked by a black cross). 
The latter is only slightly offset from that reported by \citetalias{2017ApJ...845..140Y} based on 6.0 GHz observations combining A, BnA, and B arrays (marked as the yellow cross in Figure \ref{fig:combined}), which can be possibly due to differences in array configuration, flux variation, and/or phase-referencing weather.
The measured peak (integrated) flux density is a factor of $\sim$2.5 ($\sim$3) higher than that measured by the VLA B-configuration 6 GHz observations taken in 2012.

\section{Discussion and conclusion}
\label{sec:dis}

In this work, we have presented VLBI observations of M31* combining the VLBA, Tianma-65m, and Shanghai-25m radio telescopes in four epochs in 2016, primarily at a central frequency of 4.9 GHz. 
Despite the unprecedented simultaneous resolution and sensitivity achieved, these VLBI observations do not detect the expected radio emission from M31*, which has been firmly detected by VLA observations since the 1990s. 

We consider two physical possibilities for this non-detection. 
The first possibility is that M31* experienced a temporary decrease ($\gtrsim$40\%) in its radio flux during our 2016 VLBI observations. 
As accretion-powered sources, LLAGNs are known to exhibit flux variation at nearly all wavelengths and on timescales from hours to years \citep{2008ARA&A..46..475H}.
Specifically for M31*, radio variability at 8.4 GHz was first noticed by \citet{1993ApJ...417L..61C}, which was in fact taken as a strong argument for the radio emission coming from an LLAGN.
\citetalias{2017ApJ...845..140Y} found significant (up to a fractional amplitude of 70\%) variability on timescales from days to months in the VLA 6 GHz observations between 2011--2012.
Moreover, \citetalias{2017ApJ...845..140Y} noticed that the mean flux density of M31* between 2011--2012 is $\sim$50\% lower than that measured from historical VLA observations between 2002--2005, which was $\rm 60.0 \pm10.0\  \mu Jy\ beam^{-1}$ at 5 GHz \citep{2010ApJ...710..755G}. 
Therefore, either a continued fading of M31* till 2016 or a sudden drop of flux in 2016 might explain the VLBI  non-detection. 
However, both situations seem rather unlikely, in view of the 2017 VLA detection of M31* at 6.0 GHz, which exhibited a flux density comparable to its highest level recorded during 2002--2005,
although the lack of sensitive 6.0 GHz observations between 2013--2016 prevents us from completely ruling out
these two possibilities.
A related possibility is short-term (i.e., daily) variability.  
The most significant such variability was recorded around Dec. 30th, 2012. On that day, the 6 GHz flux density of M31* was only $\rm 27.3 \pm3.7\  \mu Jy$, having decreased by  $\sim 60\%$ compared to the flux density of $\rm 65.9 \pm5.1\  \mu Jy$ observed a week before (\citetalias{2017ApJ...845..140Y}). However, it is rather unlikely that all four epochs of our VLBI observations were coincident with a temporary low-flux level. Moreover, because the light travel time is about 1/3 pc per year, whose projected distance is about 83 mas at the distance of M31, it is unlikely that M31* completely changed the $\sim \rm 1\arcsec$/$\sim$ 4 pc feature (approximately equal to the 2017 VLA synthesized beam size) between 2016 VLBI and 2017 VLA epochs. 
It is also unlikely that foreground absorption or scattering affects the observed flux over the radio frequencies, given the low line-of-sight column density toward M31* \citep{2011ApJ...728L..10L}.
Therefore, we disfavor flux variability, intrinsic or extrinsic, as the cause of the VLBI non-detection.  

\begin{figure}[h]
\centering
\includegraphics[width=0.8\textwidth]{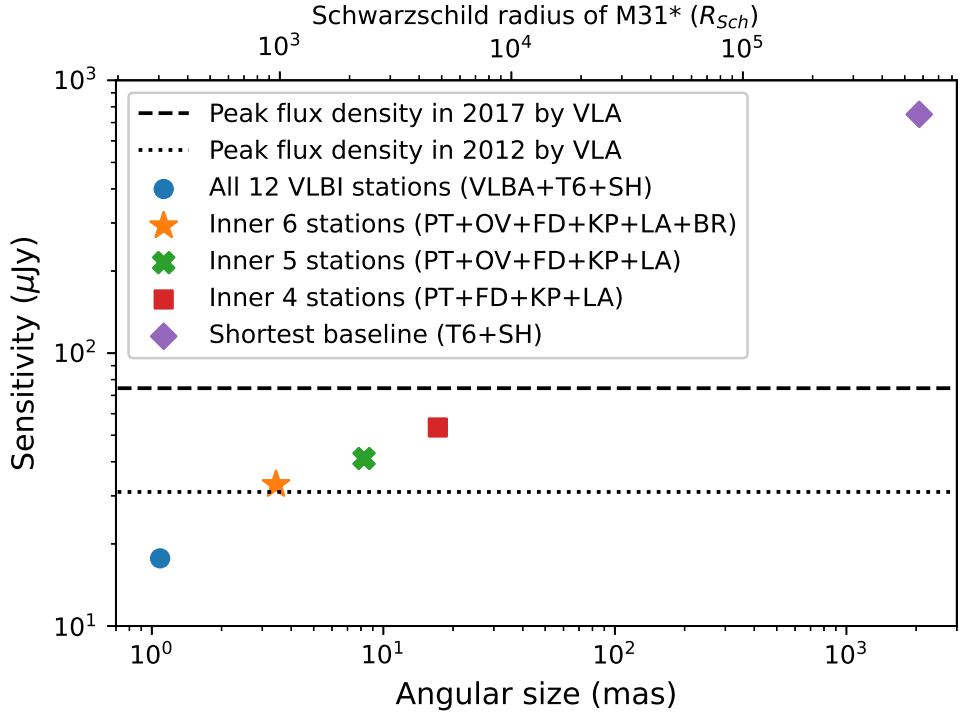}
\caption{\label{fig:angularsize} The 3$\sigma$ limiting flux density  
as a function of angular resolution or equivalent source angular size. Different symbols represent different combinations of stations, which have different angular resolutions and thus different sensitivity for an extended source of the corresponding size. 
The dotted and dashed lines denote an ideal point source at the peak flux density of M31*, as detected by VLA in 2012 and 2017, respectively. The upper horizontal axis has a scale corresponding to the Schwarzschild radius of M31*.} 
\end{figure}

A second possibility is that the compact radio source seen by VLA is intrinsically extended, such that it is largely resolved out by the VLBI observations, which have an angular resolution about three orders of magnitude higher than that of the VLA observations. 
The discrepancy between VLBI and VLA resolutions causes a large difference in their surface brightness sensitivity. For the VLBI observations (Combine\_BCD), the surface brightness sensitivity is $\rm 50\ mJy~arcsec^{-2}$, and for the VLA observation, it is $\rm 3\times 10^{-3}\ mJy~arcsec^{-2}$. Arrays  with low surface brightness sensitivity could not detect an extended source.
To quantify this possibility that the radio emission of M31* is an extended source, in Figure \ref{fig:angularsize}, we show the 3$\sigma$ limiting flux density that can be probed by different combinations of stations, as a function of source angular size corresponding to the highest angular resolution afforded by the combined stations.
Here the visibility of Combined\_BCD is adopted.
Representative combinations range from T6+SH, which has the shortest baseline, to selected VLBA stations with baselines between 200--2400 km, to all stations combined that has the longest baseline. 
For an ideal point source, the measurable flux density remains the same under different angular resolution. For an extended emission, because interferometers act as spatial filters \citep{Thompson2017}, it would easily be resolved-out and yield low flux density on long baselines. Therefore, our VLBI non-detection can rule out the possibility that M31* is a point-like source (blue circle in Figure \ref{fig:angularsize}), if it were at the same peak flux density level as the VLA observation in 2012 (horizontal dotted line) or 2017 (horizontal dashed line).The 1$\sigma$ sensitivity of the inner four stations combining PT, FD, KP, and LA (3$\sigma$ sensitivity denoted as the red square in Figure \ref{fig:angularsize}) is 17.8 $\rm \mu Jy$, which is about a quarter of the peak flux density of VLA measurement in 2017. The non-detection of the inner four antennas combination suggests that the M31* flux of 2016 was below 2017 or that M31* emission region was significantly extended beyond the angular size of the four antennas combination that is $\sim$ 17 mas or $\sim 4800~R_{\rm Sch}$. 
Similar arguments can also be applied to the 3$\sigma$ sensitivity of the inner six stations (the orange star) and inner five stations (green cross). Therefore, the M31* emission is extended and definitely more than $\sim$1 mas or $\sim 280~R_{\rm Sch}$ (all stations, blue circle), since M31* emission did not switch off as discussed in the first possibility.
Compared to the symbols below the dashed line, the 3$\sigma$ sensitivity of the shortest baseline denoted by the purple diamond is significantly worse, which is consistent with the fact that we did not detect any signal by the shortest baseline.

It is interesting to compare the above inferred size of M31* with that of Sgr A*. At 5 GHz with an arcsecond resolution by VLA, Sgr A* has a flux density of $\sim$ 0.6 Jy and a monochromatic luminosity of $\sim$ $\rm 10^{32}\ erg\ s^{-1}$ \citep{1998ApJ...499..731F,2001ApJ...547L..29Z}.
It is well known that the apparent size of Sgr A* is strongly affected by severe foreground interstellar scattering, which is approximately dependent on the square of observed wavelength \citep{2005Natur.438...62S,2006ApJ...648L.127B,2008Natur.455...78D}. 
At 5 GHz, Sgr A* appears as an unresolved source with a size of $\sim \rm 4 \times10^{3}$ times its Schwarzschild radius \citep{2004Sci...304..704B,2005Natur.438...62S,2009A&A...496...77F}. The intrinsic size of Sgr A*, which roughly scales with wavelength, can be estimated to be $\sim240~R_{\rm Sch}$ (with large uncertainty) at 5 GHz, based on the de-scattered measurements at shorter wavelengths given by \citet{2018ApJ...865..104J}.
Due to its high Galactic latitude, no significant scattering screen has been reported for M31, which means that M31* is probably free of foreground scattering. Thus, given the measured limiting size, $\sim 280~R_{\rm Sch}$, the intrinsic size of M31* is comparable to or maybe even larger than that of Sgr A*.

In terms of the radio flux distribution, a case similar to M31* was encountered in the LLAGN of NGC\,404, a nearby lenticular galaxy. 
Both VLA and Westerbork
Synthesis Radio Telescope observations detected a bright 1.4 GHz/L-band nuclear source with a negative spectral index in this galaxy at arcsec-scales \citep{2012ApJ...753..103N,2014ApJ...791....2P}, whereas a sensitive European VLBI Network (EVN) observation found no significant signal at the same frequency using a much higher angular resolution of $\sim$10 mas \citep{2014ApJ...791....2P}. At higher frequencies (15 GHz/Ku-band) observed by the VLA, this nuclear source has a centrally peaked and extended morphology, interpreted as a radio outflow associated with a confined jet  \citep{2017ApJ...845...50N}. 
More recently, the Fundamental Reference AGN Monitoring Experiment (FRAMEx) carried out an X-ray/radio observing campaign on a sample of hard X-ray-selected local AGNs \citep{2021ApJ...906...88F,2022ApJ...936...76S}. In about half of the sample AGNs observed by high-resolution VLBA observations, no significant nuclear source was detected, which is contrary to the VLA detection at lower angular resolution.
This difference suggests that the radio emission detected by the VLA may be too extended to be detected on VLBI baselines \citep{2021ApJ...906...88F}.

Physically, the inferred extended radio emission from M31* and other LLAGNs might be dominated by a relativistic jet or a non-relativistic wind/outflow driven by the hot accretion flow onto the weakly accreting SMBH. In the former case, which is widely considered for LLAGNs, the radio emission arises from synchrotron radiation of the magnetized, relativistic particles carried by the jet.
In the latter case, 
an energetic wind is a generic prediction by the theories and numerical simulations of hot accretion flows \citep{1999MNRAS.303L...1B,2012MNRAS.426.3241N,2012ApJ...761..130Y,2015ApJ...804..101Y}. 
Direct observational evidence for such a wind was recently found in M81* \citep{2021NatAs...5..928S}, a prototype LLAGN, and the LLAGN in NGC\,7213 \citep{2022ApJ...926..209S}.
Launching from the coronal region of the hot accretion flow,
the wind has a wide opening angle, which is different from a jet that is highly collimated. 
Although the wind itself is non-relativistic or sub-relativistic, it is expected that magnetic re-connection can frequently take place at the interface between the hot accretion flow and the wind \citep{2009MNRAS.395.2183Y}, giving birth to relativistic electrons.
Part of these relativistic electrons may be entrained by the wind, producing radio synchrotron emission downstream.
In addition, the outward propagating wind, with a typical velocity $\gtrsim 1000\rm~km~s^{-1}$ \citep{2021NatAs...5..928S}, can drive strong shocks into the circumnuclear medium, which in turn produce relativistic electrons and radio synchrotron emission.
This may explain the parsec-scale ``plumes'' detected in the VLA C-band image (Figure~\ref{fig:combined}b).

In summary, we have performed 5 GHz VLBI observations on M31* in 2016 with a resolution of a milli-arcsecond and sensitivity of $\rm 5.6\ \mu Jy~beam^{-1}$, but did not detect any significant signal. This non-detection, together with the successful VLA detections in 2012 and 2017, supports that M31* has an intrinsic size over $\sim$300 times the Schwarzschild radius at 5 GHz. A hot wind could be the origin of this extended radio emission, but the exact physical extent and radio flux density of both the jet-induced and wind-induced synchrotron emission are uncertain. To continue investigating the immediate surroundings of M31* in the radio regime, we have applied for new VLA observations for M31*, hoping to reveal more details of its central parsec region. Future multi-wavelength, high-resolution observations of the central parsecs of M31 hold promise for providing strong constraints on the physical properties of this putative wind. The next generation new sensitive and high angular resolution telescopes, such as the ngVLA \citep{2018ApJ...859...23N} and the Square Kilometre Array \citep{2009IEEEP..97.1482D}, with the long baseline extension will be crucial.

\begin{acknowledgments}
S.P. and Z.L. acknowledge support by National Natural Science Foundation of China (grants 11873028, 11473010) and the National Key Research and Development Program of China (2017YFA0402703). S.P. thanks Dr. Ru-sen Lu for help with data issue.
The National Radio Astronomy Observatory is a facility of the National Science Foundation operated under cooperative agreement by Associated Universities, Inc.

\end{acknowledgments}

\facilities{VLBA, VLA, Tianma/SHAO}

\software{AIPS \citep{1996ASPC..101...37V}, APLpy \citep{2012ascl.soft08017R},  CASA \citep{2022PASP..134k4501C}, Difmap \citep{1994BAAS...26..987S}.}

\bibliography{M31VLBI}{}
\bibliographystyle{aasjournal}


\listofchanges

\end{document}